# THE MASS-DEPENDENT CLUSTERING HISTORY OF $K$-SELECTED GALAXIES AT $Z < 4$ IN THE SXDS/UDS FIELD.

Junko Furusawa[1], and Kazuhiro Sekiguchi[1,2], Tadafumi Takata[1,2], Hisanori Furusawa[1], Kazuhiro Shimasaku[3], Chris Simpson[4], and Masayuki Akiyama[5]
*Draft version January 7, 2011*

ABSTRACT

We investigate mass-dependent galaxy evolution based on a large sample of (more than 50,000) $K$-band selected galaxies in a multi-wavelength catalog of the Subaru/XMM-Newton Deep Survey (SXDS) and the UKIRT Infrared Deep Sky Survey (UKIDSS)/Ultra Deep Survey (UDS). We employ the optical to near-infrared photometry to determine photometric redshifts of these galaxies. Then, we estimate the stellar mass of our sample galaxies using a standard fitting procedure as we used for estimation of the photometric redshift. From the sample galaxies, we obtain the stellar mass function of galaxies and the cosmic stellar mass density (SMD) up to $z \sim 4$. Our results are consistent with previous studies and we find a considerable number of low-mass galaxies ($M_* \sim 10^{10.5}$) at the redshift range $3 < z < 4$.

By combining stellar masses and spatial distributions of galaxies derived from a large number of galaxies in the contiguous wide and deep field, we examine properties of the mass-dependent clustering of galaxies. The correlation functions of our sample galaxies show clear evolution and they connect to that in the local universe consistently. Also, we find that the massive galaxies show strong clustering throughout our studied redshift range.

The correlation length of massive galaxies rapidly decreases from $z = 4$ to 2. The mass of dark haloes (MDHs) hosting the intermediate-mass value galaxies changes from high ($10^{14} M_\odot$) to low ($10^{13} M_\odot$) with decreasing redshift at around $z \sim 2$. We also find some high mass density regions of massive galaxies at $1.4 \leq z < 2.5$ in our sample. These concentrations of massive galaxies may be candidate progenitors of the present-day clusters of galaxies. At this redshift range, massive star forming galaxies are the dominant population making up the structures and the passively evolving galaxies show stronger clustering and they may have formed earlier than those star forming galaxies.

*Subject headings:* galaxies: evolution, galaxies: formation, galaxies: luminosity function, mass function, large-scale structure of Universe

## 1. INTRODUCTION

The Cold Dark Matter (CDM) scenario describes that the Universe is dominated by the cold dark matter and the formation of galaxies is governed by the dark matter. In the CDM universe, small structures are formed by the density fluctuations at the early epoch of the Universe and galaxies are formed by colliding and merging among those objects. Therefore, understanding the process of galaxy formation and those structures, which might have a relationship with the initial distribution of the dark matter, is important for studying the history of the Universe. For this reason, it is essential to uncover the history of galaxy formation and evolution from the early Universe. In particular, it is known that the most of the present-day massive galaxies are classified as ellipticals and a large fraction (74%) of stars in the local Universe are contained in them (Fukugita et al. 1998). These massive elliptical galaxies are usually located at the center of the clusters of galaxies and they may be formed in the massive dark matter haloes. Therefore, studying the history of mass assembly and the clustering properties of such a galaxies are important to understand the history of the Universe.

To make sure that we have a statistically meaningful sample for studying the evolution of galaxies, it is necessary to examine a large number of galaxies from high to low redshift. Previously, a variety of methods using the galaxy colors have been used to select samples of the high redshift (high-$z$) galaxies. Those methods allow us to investigate some specific galaxy populations which are efficiently detected in limited redshift intervals. For example, Extremely Red Objects (EROs) are selected by the color criterion of $R$-$K$ or $I$-$K$, which picks up the red galaxies at $z > 1$ (Elston et al. 1998; Thompson et al. 1999; Cimatti et al. 2002; McCarthy 2004). Distant Red Galaxies (DRGs) are the sample of the red galaxies at $z=2$ to 4 selected using $J$-$K$ color (Franx et al. 2003). These selection methods are intended for investigating progenitors of massive elliptical galaxies. The $BzK$ galaxies ($BzK$s) are high-$z$ galaxies at $1.4 \lesssim z \lesssim 2.5$ which are selected based on $B$-$z$ and $z$-$K$ colors (Daddi et al. 2004). They are further divided

Electronic address: furusawa.junko@nao.ac.jp
[1] National Astronomical Observatory of Japan, 2-21-1 Osawa, Mitaka, Tokyo 181-8588, Japan
[2] Department of Astronomical Science, School of Physical Science, The Graduate University for Advanced Studies, Shonan Village, Hayama, Kanagawa 240-0193, JAPAN
[3] Department of Astronomy, School of Science, the University of Tokyo, 7-3-1 Hongo, Bunkyo-ku, Tokyo 113-0033, Japan
[4] Astrophysics Research Institute, Liverpool John Moores University, Twelve Quays House, Egerton Wharf, Birkenhead CH41 1LD, UK
[5] Department of Astronomy, Faculty of Science, Tohoku University, 6-3, Aramaki Aoba, Aoba-ku, Sendai, Miyagi 980-8578, Japan

into star forming $BzK$ galaxies (s$BzK$s) and passively evolving $BzK$ galaxies (p$BzK$s). These two categories represent different histories of the star formation activities. The Lyman-break technique (Steidel et al. 1996) is a method of selecting star-forming galaxies using the redshifted Lyman break (912 Å at the rest frame) in the Spectral Energy Distribution (SED) of the galaxies. The galaxies selected by this technique are called Lyman Break Galaxies (LBGs) (Giavalisco 2002, and references therein).

To date, there are numbers of investigations on the spatial mass assembly and on the clustering of these color-selected galaxy populations (e.g. Ouchi et al. 2005 and Yoshida et al. 2008 for LBGs; Grazian et al. 2006b, Foucaud et al. 2007, Quadri et al. 2008 and Tinker et al. 2010 for DRGs; Kong et al. 2006 and Quadri et al. 2007 for EROs; Kong et al. 2006, Hayashi et al., and Blanc et al. 2008 for $BzK$s). Some of these studies suggest that at least a part of these populations become the present-day massive elliptical galaxies (e.g. Miyazaki et al. 2003, Kong et al. 2006, and Hayashi et al. 2007). However, the redshift ranges studied using these color-selected galaxy populations are too wide to examine details of the evolution of galaxies at their corresponding epochs. Also, those color-selected populations represent only a part of the whole galaxy populations at their corresponding redshift ranges. Furthermore, it is known that there may be significant contaminations in these color-selected galaxy samples (Reddy et al. 2005; Grazian et al. 2007). Thus, by using only the color selection method, it is difficult investigating detailed views of the galaxy evolution. Spectroscopic data or the photometric redshift (photo-$z$) technique with multi-band photometric data, such as used in our present study, is needed for determining more accurate redshifts of such galaxies.

Investigations of stellar mass and luminosity of various types of galaxies from high to low redshift are required for statistically understanding the history of the galaxy evolution. Specifically, it is useful to study changes in those physical parameters at different redshifts. For better understanding the evolution of the stellar mass function of galaxies up to higher redshift and to smaller stellar mass, a deeper observation data is needed. Recently, such studies for the $K$-selected galaxies have been performed up to $z$ =4 to 5 [Drory et al. 2005 ($z < 5$); Fontana et al. 2006 ($z < 4$); Pozzetti et al. 2007 ($z < 2.5$); Pérez-González 2008 ($z < 4$); Drory et al. 2009 ($z < 1.0$; spectroscopic redshifts)]. For the IRAC 3.6$\mu m$-selected galaxies, Ilbert et al. (2010) investigated the evolution of the galaxy stellar mass function at $0.2 < z < 2$. In addition to investigate the stellar mass and luminosity in various different redshifts, it is useful to study the clustering property of those galaxies. So far, the galaxy clustering as a function of the stellar mass of galaxies, including the field galaxies, was derived only for low redshift samples [e.g. Meneux et al. 2009 ($z \sim 1.0$)]. For understanding the clustering property at high-$z$, Daddi et al. (2003) investigated the $K$-band luminosity-dependent clustering of galaxies at $z \leq 4$. However, the field area studied by them was too small (about 4$arcmin^2$) to statistically describe the evolution of the massive galaxies.

It is known that the correlation length ($r_0$) of the local clusters of galaxies is $\sim 15 - 20 Mpc$ (Bahcall et al. 2003). To avoid the field-to-field variation due to the cosmic large-scale structures, it is desirable to cover two or more fields of $20 \times 20 Mpc$ in size at high-$z$ (e.g., at least 0.30$deg^2$ at $z \leq 1.4$). Of course, it is not easy to obtain a contiguous wide and deep image, which demands a large observation time especially in the infrared wavelengths. Nevertheless, several attempts had been taken and the literatures suggest as follows: (1) The evolution of more massive galaxies is active at the earlier epoch of the Universe and these galaxies are older (Pérez-González et al. 2008). (2) The peak of star formation density was at some time between $z = 3$ and $z = 1$ (Steidel et al. 1999, Chapman et al. 2005, and Le Floc'h et al. 2005). (3) The red sequences of massive field galaxies are detected at $z < 2.3$ (Kriek et al. 2008) and the merger density peak of massive galaxies is estimated to be at $z \sim 1.2$ (Ryan et al. 2008). (4) At $z < 1$, the star formation rate of the low-mass galaxies is high (Bell et al. 2007).

Unfortunately, so far, there is no statistically sound observational result which allows us to comprehensively examine the history of galaxy evolution and formation covering the sufficiently wide stellar mass and redshift ranges. In this paper, we present the stellar mass properties and the clustering properties of $K$-selected galaxies at redshift $0.6 < z < 4.0$ in the SXDS (Subaru/XMM-Newton Deep Survey: Sekiguchi et al, 2004, Furusawa et al. 2008)/UDS (Ultra Deep Survey) of the United Kingdom Infrared Telescope Infrared Deep Sky Survey (UKIDSS: Lawrence et al. 2007) field. A contiguous deep ($K_{AB} <= 23.5$) and wide field (0.63$deg^2$) allows us to discuss the formation and evolution of galaxies up to $z = 4$ in details. Using multi-band images we are able to estimate the photometric redshifts of galaxies, which are more accurate than the redshift range derived only by using color-selected techniques, and the stellar masses of the galaxies. By paying attention to the stellar mass, we will discuss statistically the evolution of galaxies and their the clustering history.

In section 2 we will describe the data we use, and explain how we estimate the photometric redshift and the stellar masses, and we will present clustering analysis in section 3 and 4. In section 5 we will discuss the evolution of stellar mass and clustering of galaxies based on our results, and give the summary in section 6. The relevant conversion between Vega and AB magnitudes for this paper is $K_{AB} = K_{Vega} + 1.89$. We use a cosmology with $\Omega_\Lambda = 0.7$, $\Omega_m = 0.3$, and $H_0 = 70 \mathrm{km s^{-1} Mpc^{-1}}$ ($h_{70} = H_0/70$) throughout this paper, except for the correlation length, which is given using $h = H_0/100 \mathrm{km s^{-1} Mpc^{-1}}$, for the sake of comparison with previous results.

## 2. DATA

In this study, we use an optical data set obtained by the SXDS (Furusawa et al. 2008) and a near-infrared data set obtained by the UKIDSS/UDS (Warren et al. 2007). The SXDS field is centered at RA(J2000) = $2^h 18^m 00^s$, DEC(J2000) = -5°00'00" and it covers a contiguous field of $\sim 1$ square degree (Figure 1).

### 2.1. *The optical data*

An optical imaging survey part of the SXDS was obtained by Subaru Prime Focus Camera (Suprime-Cam, here after) (Miyazaki et al. 2002) on the Subaru Tele-scope at MaunaKea, Hawaii. A total of five Suprime-Cam pointings are laid out as cross shape so that each of the North-South and the East-West direction has an extent of ∼1.3 degree (Figure 1) and it covers ∼1.22 square degrees. The $B, V, R_c, i'$, and $z'$ imaging observations are carried out to the depths of $B = 28.4$, $V = 27.8$, $Rc = 27.7$, $i' = 27.7$, and $z' = 26.6$ (AB, 3 sigma and $\phi = 2"$) with a typical seeing of ∼0.8 arcsec FWHM (Furusawa et al. 2008). With an exception of the $z'$-band, the depths of these optical imaging data are comparable to those reached by the ACS in the two 160 arcmin² GOODS fields. The SXDS observations were conducted in a period from September 2002 to September 2005.

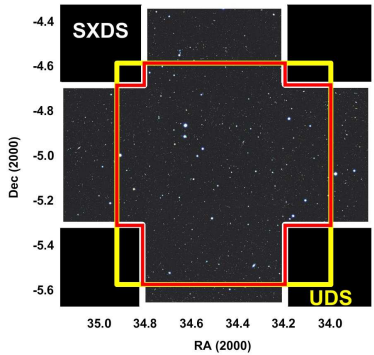

FIG. 1.— Composite image of SXDF. $B$-, $R_c$-, and $i'$-band are assigned to RGB colors. An area defined by the red line is the overlap region of the SXDS (white cross) and the UKIDSS/UDS (yellow square).

The current version of the $i'$-band data consists of nearly a million objects (SXDS DR1: Furusawa et al. 2008). The optical object catalogs deduced from the SXDS are open to public access through the official web site (http://www.naoj.org/Science/SubaruProject/SXDS/). These SXDS photometric catalogs are one of the deepest and the widest multicolor data set available to date. The data set has been successfully used, with various data in other wavebands on the same field, to study distant galaxies (e.g., Hartley et al. 2010).

### 2.2. The near-infrared data

The UKIDSS/UDS (Ultra Deep Survey) is a deep near-infrared imaging survey, which aims to cover a ∼0.77 deg² field to aim the final 5σ depths (AB mag, 2" diameter aperture) of $J$=26.5, $H$=25.5 and $K$=25 (see http://www.ukidss.org for more information). The UKIDSS/UDS is one of the widest-area near-infrared surveys of this depth ever performed. The UKIDSS/UDS survey was initiated in September 2005 and is currently on going.

Our study uses the UKIDSS/UDS Data Release 1 (DR1: Warren et al. 2007) data set, which consists of $J$- and $K$-band imaging of the entire UDS field to the 5σ depths of $J_{Vega}$=22.61 and $K_{Vega}$=21.55. In terms of the $K$-band depth, the UKIDSS/UDS DR1 data are as deep as the dataset in the GOODS-CDFS field obtained by VLT-ISAAC (Moorwood et al. 1998). The UKIDSS/UDS is the largest contiguous near-infrared image of this level of depth so far available.

### 2.3. The optical-infrared catalog

In this work, we use the SXDS and UKIDSS/UDS DR1 source matching catalog, which we constructed. A 7-band multi-color catalog is created by combining dataset from the SXDS and the UKIDSS/UDS using SExtractor (Bertin & Arnouts, 1996) in the following manner: First, the UDS $K$-selected catalog is matched to the SXDS $i'$-band catalog, using 1"-radius margins for the cross identification, combining the optical 5-band magnitudes and the infrared 2-band magnitudes. Then, to obtain reliable aperture magnitudes in multiple bands, astrometric information of the optical SXDS images is re-calibrated by using the $i'$-band images and the $K$-band images so that optical and near-infrared positions should be identical for all the objects. Finally, the magnitudes of SXDS objects in the optical bands are measured again on the SXDS images at the positions of the $K$-selected UDS objects with circular apertures. In this study, we use magnitudes and colors of objects measured in the 2"-diameter apertures for studies on the number counts and detection completeness of objects, and selection of galaxy populations performed in this section. MAG_AUTO is used for photometric redshifts and studies on the stellar masses of galaxies in Section 3 to 5. The magnitude errors are estimated based on the Poisson noise of the sky-background fluctuations.

We adopt $5\sigma_{sky}$ limiting magnitude in $K$ band $K_{Vega} = 21.55$ to make a magnitude-limited sample. Star-galaxy separation is made based on FWHM of objects and the $B$-$z'$ and $z'$-$K$ colors (Daddi et al. 2004). There are a total of 60,175 galaxies identified in a 2,279 arcmin² area. We use the $K$-selected sample for which photometric data are available in at least 5 bands in addition to the $K$ band, in order to maintain higher quality of the template fitting analysis of photometric redshift and estimation of stellar mass of galaxies. This criterion is satisfied by 55,174 galaxies (hereafter we call them the SXDS/UDS sources). Figure 2 shows a comparison of the $K$-band number counts of galaxies in our sample with the counts from the literature. The number counts provide a statistical probe for both the spatial distribution and evolution of different populations of galaxies (Figure 3). For this reason, we also compare the number counts of various populations, namely, sBzKs, pBzKs (star-forming and passive $BzK$s: Kong et al. 2006), DRGs, and EROs from our data with those from the other fields. As shown in Figure 2, our number counts are in good agreement with those of previous surveys. However, our sBzKs number count is higher than those in previous surveys by up to a factor of 3-4 at $K_{Vega} = 18.5$ (Figure 3). The limiting magnitude and area of these comparison fields are listed in Table 1. Due to the effect of the bandpass difference, our $BzK$ selection and the original selection criteria by Daddi et al. (2004) are about 15% different. Detailed information about this difference has been described by Hayashi et al. (2007).

### 3. PHOTOMETRIC REDSHIFT AND STELLAR MASS ESTIMATION

We determine photometric redshifts of the SXDS/UDS sources using the Hyperz code (Bolzonella et al. 2000). Since the accuracy of photo-$z$ depends on the model SEDs and the band set used in template fitting, we carefully inspect our photo-$z$ obtained with the photometric data in the $B, V, R_c, i', z', J, K$ bands for the two different model codes, i.e., PEGASE.2 (Fioc & Rocca-Volmerange 1997) and BC03 (Bruzual & Charlot 2003) if there was any systematic effect in our photo-$z$s. By comparing the photo-$z$s of our data with their spectroscopic redshifts (spec-$z$s), we find that the number of outliers estimated using PEGASE.2 models is smaller than that using BC03 models. This agrees with Grazian et al. (2006a), who described that the photo-$z$ obtained from PEGASE.2 models agrees better with spec-$z$ than those from BC03. We also investigate difference in photo-$z$ of galaxies for the three different initial mass functions (IMFs); Rana & Basu (1992; RB92), Salpeter (Salpeter 1955), and Chabrier (Chabrier 2003). We found no significant difference among these three IMFs. However, when we compare photo-$z$ with spec-$z$, the RB92 produces a smaller number of outliers photo-$z$ which have large errors than the other IMFs. Therefore, in this paper, we adopt the results using the PEGASE.2 code and the RB92 IMF with lower and upper cutoff masses of 0.1 $M_\odot$, and 120 $M_\odot$, respectively.

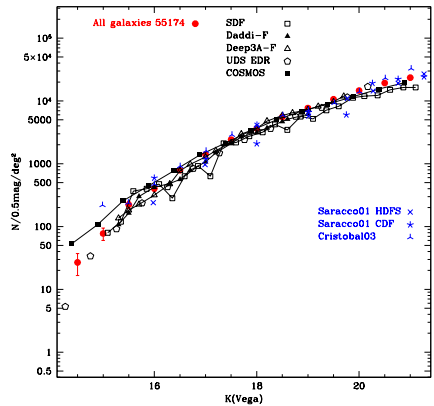

FIG. 2.— Number counts of galaxies in the SXDS/UDS field. Our data points (filled red circles) are compared with the number counts in the previous wide-field surveys. No correction for the difference between $K$ and $K_s$ magnitudes are made. The different symbols are derived from different surveys: cross – HDFS (Saracco et al. 2001), five-pointed stars – CDF (Saracco et al. 2001), three-pointed stars – Cristobal et al. (2003). We also compare our result with those from the studies on galaxy populations in the near-infrared wavebands: open squares – Subaru Deep Field (SDF; Hayashi et al. 2007), filed and open triangles – Daddi-F and Deep3A-F, respectively (Kong et al. 2006), open pentagons – UDS EDR (Simpson et al. 2006), and filled squares – COSMOS (McCracken et al. 2010).

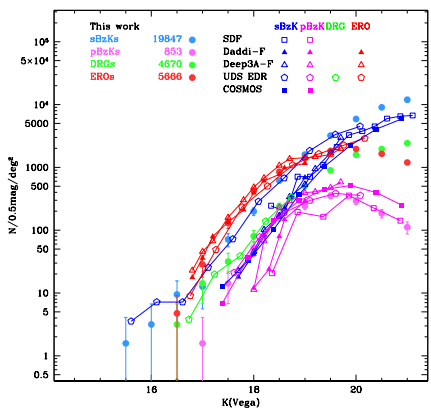

FIG. 3.— Number counts of galaxies for various populations in the SXDS/UDS field. No correction for the difference between $K$ and $K_s$ magnitudes are made. Symbols in different colors represent the number counts of different galaxy populations as a function of $K_{Vega}$ magnitude: blue – sBzK's, magenta –pBzK's, green – DRG's, and red – EOR's. We also compare our results with literature data: open squares – SDF (Hayashi et al. 2007), filled and open triangles – Daddi-F and Deep 3A-F (Kong et al. 2006), open pentagons – UDS EDR (ERO's from Simpson et al. 2006; sBzK's and pBzK's from Lane et al. 2007; DRG's from Foucaud et al. 2003), filled squares – COSMOS (McCracken et al. 2010).

| Field | $K_{lim}(AB)$ | Area (arcmin$^2$) |
|---|---|---|
| SXDS/UDS-F (This work) | 23.5[1] | 2279 |
| SDF[a] | 23.5[2] | 180 |
| Daddi-F[b] | 21.5[1] | 600 |
| Deep3A-F[c] | 22.7[1] | 320 |
| GOODS-South-F[d] | 23.5[3] | 143 |
| COSMOS-F[e] | 23.0[4] | 6804 |

TABLE 1
A list of AB magnitude and area coverage (arcmin$^2$) of deep optical-near infrared survey fields: (a) SDF (the Subaru Deep Field) from Hayashi et al. 2007, (b) Daddi-F (Daddi Field) and (c) Deep3A-F (Deep3A Field) from Kong et al. 2006, (d) GOODS-South-F (the Great Observatories Origins Deep Survey-South Field) from Fontana et al. 2006 and (e) COSMOS-F (the Cosmic Evolution Survey Field) from McCracken et al. 2010. (1) AB magnitude, $5\sigma$, $\phi=2"$. (2) We convert the total magnitude listed in Hayashi et al. (2007) into 2"-aperture magnitude by using a relation of mag(total)-mag($\phi=2"$) = -0.3 mag. (3) A typical magnitude limit in the $K_s$ band for most of the sample derived from Fontana et al. (2006). (4) The $K_s$ limiting magnitude by the completeness fraction (70% for disks and 90% for stars and bulges) from McCracken et al. (2010).

The parameters used to generate the SED synthetic models, such as the time scale for gas infall on a galaxy ($\tau_{infall}$), the time scale for exponentially decreasing star formation rate (SFR) ($\tau$ : SFR(t) $\propto$ exp(-t/$\tau$)) , the age of a galactic wind ($\tau_{wind}$), and the geometrical configuration to compute internal extinction are listed in Grazian et al. (2006a). In addition, we adopt two types of SFRs (i.e., SFR with instantaneous burst and constant SFR). Ages of the instantaneous burst model are set to 2 Gyr or less. For the constant SFR models, which represent passively evolving galaxies, we use truncation ages to stop star formation at 100, 300, 600, and 1,000 Myr. The galaxy ages of those models are set larger than truncation ages.

These parameters used in our study are similar to those used in Grazian et al. (2006a), which demonstrated successfully the accurate estimation of photo-$z$ of galaxies. In our study of the model SED templates, we find that inclusion of the younger age (under 20 Myr) model galaxy introduces a systematic bias around photometric redshift ($z_{photo}$) = 3 and it leads to a larger fraction of massive galaxies which are possibly misidentified, especially at around $z = 1.5$ and higher redshifts. Because of these reasons, the ages of the SED models in this work are limited to older than 20 Myr over the redshift range $0 \leq z \leq 6$, for which we have model SEDs.

### 3.1. Comparison of the photometric and the spectroscopic redshifts

In Figure 4, we compare our photo-$z$s of a subsample of galaxies with spectroscopic redshift measurements

(spec-$z$), available to us (Akiyama et al. 2010, Ouchi et al. 2010, and Furusawa et al., in preparation). Agreements between the photo-$z$s and the spec-$z$s are good even if galaxies with broad-emission line are included in the sample. However, caution should be taken to interpolate these good agreements at the high redshift $z > 2$, because the number of galaxies compared is relatively small, and there is only one spectroscopic sample at $z > 4$. We estimate catastrophic errors using the method used in Ilbert et al. (2009). The percentages of catastrophic failures on our data are 6% at $z \leq 2$ and 18% at $z > 2$. The accuracy of photo-$z$s at $z \sim 3$ or $z <\sim 0.6$ may be affected by lack of $U$- and $H$-band data in estimation of the photo-$z$s. However, no large systematic errors in the photo-$z$s at those redshift ranges are seen in comparison with the limited number of spectroscopic redshifts.

Further to ensure our photo-$z$ estimations, we apply our method to a 668 galaxy sample in the GOODS-MUSIC catalog (Grazian et al 2006a) to compare with. Our results for the GOODS-MUSIC sample agree very well with those shown in Grazian et al. (2006a) for the same galaxy sample. Quantitatively, in spite of the fact that we use only 6 bands ($BVizJK_s$) to obtain photo-$z$, the deviation of the photo-$z$ from spec-$z$, i.e., $\Delta z = (z_{photo} - z_{spec})$ is quite small, and is given by the Gaussian distribution with a standard deviation $\sigma = 0.06$. This deviation is about the same level as that obtained by Grazian et al. (2006a) with 14 bands, although the error distribution of the photo-$z$s may be slightly asymmetric in separate redshift ranges.

consistent with photo-$z$ and other physical parameters of the best-fit SEDs, including age, metallicity, extinction due to the internal dust etc..

We examine typical errors of the estimated stellar masses using the technique based on the reduced chi-square ($\chi^2$), which is similar to those used in Papovich et al. (2001) and Fontana et al. (2004). First, we obtain the variation in stellar mass ($M_*$) of each galaxy which is obtained in the $\chi^2$ range between $\chi^2_{bestfit}$ and $\chi^2_{bestfit}+1$. Then, the mass range thus determined is regarded as a typical error of the stellar mass estimation. We find that the typical error on the estimated stellar masses is on the order of $\pm 0.2 dex$. It also turns out that the stellar mass on average derived with the RB92 IMF is 1.5 times smaller than that with the Salpeter IMF when we use the same photo-$z$ to the two estimations.

We calculate the 95% completeness level of $M/L$ ratio using the same method by Pozzetti et al. (2009). Figure 5 shows the stellar mass as a function of redshift and the limiting stellar mass ($M_{lim}$) at the 95% completeness level derived from our data.

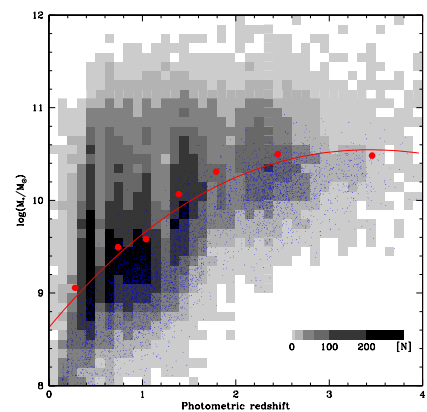

FIG. 5.— The distribution of stellar masses of the SXDS/UDS galaxies as a function of redshift. The density in each grid of the shaded gray area show the number count of galaxies which falls in the mass and redshift ranges ($\Delta z = 0.1$ and $\Delta log\{M_*/M_\odot\} = 0.1$). The blue dots superimposed show the distribution of stellar masses of the the faintest galaxies (20% of the total sample galaxies) of our sample, which contribute to the stellar mass near the magnitude limit at each redshift. The red filled circles show the limiting stellar mass $M_{lim}$ at the 95% completeness level in the $M/L$ for respective redshift ranges. The red curve in the midst of the red circles is the best-fit $M_{lim}$ against redshift.

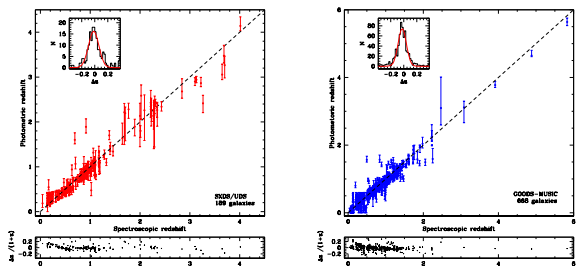

FIG. 4.— Comparison of the photo-$z$ with the spec-$z$ from the SXDS/UDS sample (left) and GOODS-MUSIC sample (right). For each point, the error bar shows the confidence interval at 90%. At the upper left of the each panel, the histogram of the scatter $\Delta z = (z_{photo} - z_{spec})$ is shown and compared with Gaussian distribution with a standard deviation (red curve). The dispersions are $\sigma = 0.08$ in the left panel and $\sigma = 0.06$ in the right panel. The scatter $\Delta z/(1 + z_{spec})$ as a function of the spec-$z$ is shown below the each panel.

### 3.2. Stellar mass estimation

We estimate stellar masses of our sample galaxies based on the template SED fitting. The stellar mass to light ratios in the $B$ band ($M/L_B$) is derived from the best-fit SED determined by the template fitting at the same time as the photo-$z$ is determined. The $M/L_B$ is converted to stellar masses of galaxies by calculating absolute $B$ magnitude of galaxies based on the best-fit SED and the photo-$z$. Since we employ exactly the same SED models both in the photo-$z$ determination and mass estimation, stellar masses of galaxies thus obtained are internally

We derive the Galaxy Stellar Mass Functions (GSMFs) using the $V_{max}$ formalism (Schmidt 1968) at different redshift ranges (Figure 6). Our results for GSMFs are fitted with a smooth function using a Schechter (1976) parametrization. We fit our GSMFs using completeness-corrected data (the stellar masses above our 95% completeness level at each redshift) points and uncertainties.

These are compared with GSMFs from previous studies in other fields, with roughly the same redshift range. In particular, our GSMFs are similar to the $K_s$-selected mass functions from GOODS-MUSIC(U-4), derived by Fontana et al. (2006), which used Spitzer data. In contrast, we see significant variance at the massive end among the results from different surveys. The GSMFs in this study and COSMOS (for $z = 1.6 - 2.0$) are system-atically lower than those by Elsner et al. (2008) using GOODS-MUSIC(U-K) data without the Spitzer/IRAC photometry. Elsner et al. (2008) found that stellar mass tends to be overestimated on average without the Spitzer photometry by the analysis using the Salpeter IMF. However, the GSMFs of Pérez-González et al. (2008), in spite of inclusion of the Spitzer data in their analysis, are more consistent with those of GOODS-MUSIC(U-K) than GOODS-MUSIC(U-4). This inconsistency may be due to the choice of template models (e.g. IMF, which we described above, and the spectral synthesis model), the difference in bands used for the estimation of stellar masses (i.e. Fontana et al. 2006; Elsner et al. 2008), and the cosmic variation (Pérez-González et al. 2008). In summary, our GSMFs are in a good agreement with the previous studies over the whole mass range up to $z \sim 4$, except for the high-mass end, where our data show systematically lower stellar masses. Our GSMFs should not be affected so much from the cosmic variation such as the difference of the number density of galaxies among small fields, since our field (2279 $arcmin^2$) is much larger than GOODS-MUSIC (143.2 $arcmin^2$), FDF (FORS Deep Field, 40 $arcmin^2$), GOODS (50 $arcmin^2$), HDF-S (257 $arcmin^2$), CDF-S (225 $arcmin^2$), and LHF (183 $arcmin^2$). In addition, our deep and wide field extends both effective redshift range and stellar mass range. As a result, we obtain more reliable GSMFs in the $K$ band than previous surveys.

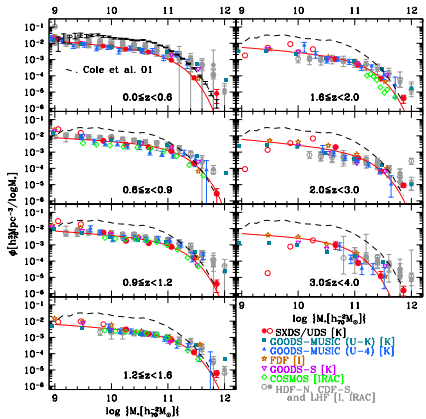

FIG. 6.— The stellar mass functions of the SXDS/UDS sample. Our results are shown by red open and filled circles within each redshift range as labeled. The filled circles show our data with the $M/L$ completeness level $> 95\%$ and the open circles show that with the completeness $< 95\%$. The red solid lines show the results with fitting by the Schechter functions using our data with the completeness larger than 95%. The black dashed lines represent the local GSMF of Cole et al. (2001). We also compare our GSMF plots and standard errors to values from other studies: Filled squares [GOODS-MUSIC ($U$-$K$), $K$-selected] from Elsner et al. (2008), filled triangles [GOODS-MUSIC (U-4), $K$-selected] from Fontana et al. (2006), open stars [FDF, $I$-selected] and open triangles [GOODS-S, $K$-selected] from Drory et al. (2005), open diamonds [COSMOS, IRAC-selected] from Ilbert et al. (2010), and open ($I$-selected) and filled (IRAC-selected) gray circles (HDF-N, CDF-S, and LHF) from Pérez-González et al. (2008). The errors from GOODS-MUSIC ($U$-$K$), FDF, GOODS-S, and COSMOS are not represented.

### 3.3. Stellar mass density

We estimate the cosmic stellar mass density (SMD) as a function of redshift from two mass ranges $M_* \geq 10^8 M_\odot$ and $M_* \geq 10^{10.86} M_\odot$ (Figure 7). For the redshift ranges which the detection completeness is low, the mass density is derived by integrating the mass function using the best-fit Schechter parameters, instead of summing up the individual masses over $V_{max}$. To evaluate our result, we compare our estimations with the stellar mass densities by other authors available in the literature. It is known that the steep faint-end slope of the Salpeter IMF below $\sim 1 M_\odot$ gives larger mass to light ratio ($M/L$) and stellar mass by a factor of $\sim 1.7 - 2$ than shallower IMF(e.g., Bell et al. 2003; Erb et al. 2006). In this paper, we refer to the values of the local stellar mass density $\rho_*$, in which the values obtained from Cole et al. (2001) and Bell et al. (2003) are divided by 1.96. From Figure 7, our results agree well with the literature reported values for each redshift range. Especially for the galaxies with $M_* \geq 10^{10.86} M_\odot$, we see that the change in SMD at $z < \sim 2$ is much more gradual than that at $z > \sim 2$.

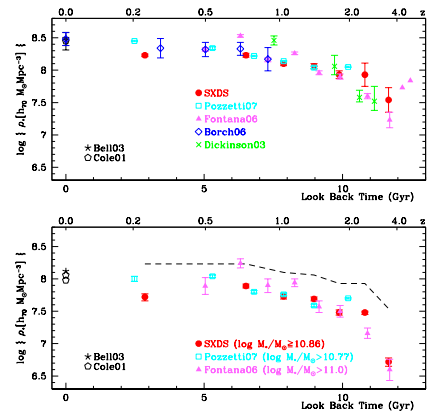

FIG. 7.— Cosmological evolution of the stellar mass density of the SXDS/UDS sources. Our estimations are plotted with red filled circles. Upper panel shows the SMD and the standard error from $\log(M_*/M_\odot) = 8$ to 13 based on the integration of the stellar mass function with a Schechter parametrization. Lower panel shows our observed SMDs and the errors computed for galaxies, whose masses are equal to or more massive than $\log(M_*/M_\odot) = 10.86$. The dashed line represents our SMD of the filled circles in the upper panel for comparison. Our results are compared with other estimates from the literature: stars from Bell et al. (2003), open pentagons from Cole et al. (2001), open squares from Pozzetti et al. (2007), triangles from Fontana et al. (2006), crosses from Dickinson et al. (2003) and diamonds from Borch et al. (2006).

## 4. CLUSTERING ANALYSIS

We study the clustering properties of the $K$-selected galaxies at high redshift in our SXDS/UDS sample by examining the distribution of galaxies with different stellar mass ranges. Here, we discuss only the galaxies with stellar mass which is larger than $\log(M_*/M_\odot) = 9.66$, for which the detection completeness is $> 70\%$ at $z = 3-4$. Figure 8 shows the projected two-dimensional distribution of the galaxies whose stellar masses are equal to or larger than $10^{11.26} M_\odot$ at $1.4 \leq z < 2.5$. We find that there are several over density regions where these massive galaxies are clustered. The combinations of the dense and sparse regions of the massive galaxies compose several possible over-dense structures, which have a projected diameter of $\sim 8$ Mpc in the comoving scale at $1.4 \leq z < 2.5$. Figure 9 shows the distributions of photo-$z$ of the galaxies with $M_* \geq 10^{10.86} M_\odot$ in three high mass-density regions and one typical mass-density region as a control field. The first region shows 2 peaks of concentration of galaxies in the photo-$z$ distribution. The second and third regions show the concentration of galaxies at $z \sim 2.35$ and $2.20$, respectively. These redshift distributions suggest the cluster candidates, which are traced by the massive galaxies.

In order to discuss the clustering properties of the galaxies quantitatively, we derive the two-point angular correlation functions (ACFs) $w(\theta)$, using the technique proposed by Landy & Szalay (1993):

$$w_{obs}(\theta) = \frac{[DD] - 2[DR] + [RR]}{[RR]}, \quad (1)$$

where $[DD]$ is the number of the data-data sample, $[DR]$ is the number of data-random sample, and $[RR]$ is the number of the random-random sample. It is known that the estimator above is slightly biased to lower values with respect to the real correlation function $w(\theta)$ ((Groth & Peebles 1977 and Daddi et al. 2000). The offset is defined as follows:

$$w(\theta) = w_{obs}(\theta) + \sigma^2, \quad (2)$$

where $\sigma^2$ is called the "integral constant" (Groth & Peebles 1977). The amplitude of the real two-point correlation function $w(\theta)$ can generally be described by a power law of the form $w(\theta) = A\,\theta^{-\delta}$. Therefore, the amplitude and the slope of the real ACF can be estimated by fitting the following function to the measured $w_{obs}$:

$$w_{obs}(\theta) = A(\theta^{-\delta} - C), \quad (3)$$

where $C = \sigma^2/A$. We fit the power low function to the ACFs of our samples. Our fitting is made at the bin centers ranging from 7.9 arcsecond to 316.2 arcsecond.

To obtain a reliable fit from each subsample, the lower limit is set to avoid a fitting problem due to nonlinear small-scale clustering and the upper limit is set much smaller than the survey field. When the angular correlation function is represented by a power law, the spatial correlation function (SCF) should also be a power law, $\xi(r) = (r/r_0)^{-\gamma}$. Here we adopt the constant value $\delta = 0.8$ and $\gamma = 1.8$ for our analysis, these values are consistent with most previous observations. The Figure 10 shows the ACFs of our samples selected with stellar mass bin ($10^{10.86} M_\odot \leq M_* < 10^{11.26} M_\odot$) in each epoch. The error bars are computed using $2 \times \delta w(\theta)$, where $\delta w = \sqrt{[(1 + w(\theta)]/[DD]}$ (Daddi et al. 2000). The $2\sigma$ errors computed in this way are equivalent to the errors obtained with the boot-strap resampling method (Baugh et al. 1996). The solid lines are the best-fit power law of Equation (3). The dotted lines represent the correlation functions from SDSS for the absolute magnitude range of $-21 \geq M_{r^*} \geq -22$ (Budavári et al. 2003). The stellar mass range from SDSS is scaled to $10^{11.0-11.5} M_\odot$, using the K-correction code and the luminosity evolution model (Li et al. 2006). This range is roughly comparable to our subsamples selected with ($10^{10.86} M_\odot \leq M_* < 10^{11.26} M_\odot$) in Figure 10. From a given angular clustering measurement, $r_0$ of galaxies for each mass range can be estimated by the Limber equation (Limber 1953) using the redshift distribution of galaxies and cosmology (e.g., Peebles 1980). Here, we use a redshift distribution based on the photo-$z$'s for the Limber transformation in each redshift bin and each stellar mass bin. For comparison with most of previous results, the correlation length is expressed in the unit of $h^{-1}$ Mpc, where $h = H_0/100 \mathrm{km} s^{-1} \mathrm{Mpc}^{-1}$. We list the results of power-law fitting of the ACFs and the estimated $r_0$ in Table 2. Although the $r_0$ values listed in Table 2 are estimated without photo-$z$ errors, we also examine the change in the $r_0$ using a top-hat redshift distribution as an extreme case. We see that the systematic errors introduced by the photo-$z$ errors are negligible, e.g., $r_0 = 22.9^{+2.3}_{-2.5}$ for $\log(M_*/M_\odot) \geq 11.26$ at $1.4 \leq z < 2.5$ (Table 2) and $r_0 = 23.9$ for the top-hat redshift distribution. On Figure 11, we present an overview of the evolution of the mass-dependent galaxy clustering. The $r_0$ lines of dark haloes (DHs) are based on the work of Mo & White (2002) and the formalism of Bardeen et al. (1986).

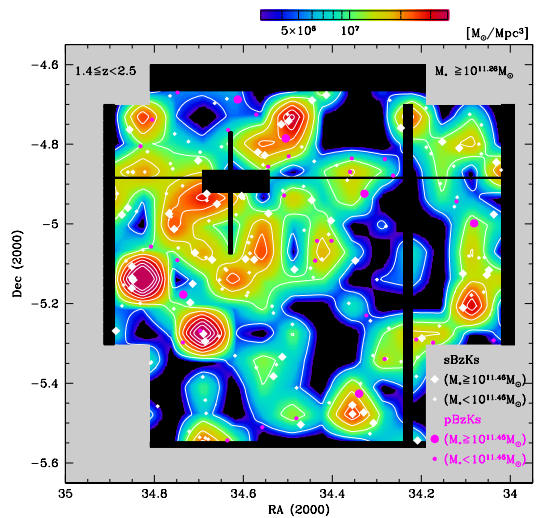

FIG. 8.— The mass density map of massive galaxies ($M_* \geq 10^{11.26} M_\odot$) at the redshift range $1.4 \leq z < 2.5$. The thick contour line shows the average ($10^{9.64} M_\odot/Mpc^3$) of the mass density map. The thin contour lines show the $\sigma$ levels from $-1$ to $5$ with respect to the average mass density. The grid size adopted here is $0.06$ deg. Sky positions of the $sBzKs$ and the $pBzKs$ are shown by filled squares and circles, respectively.

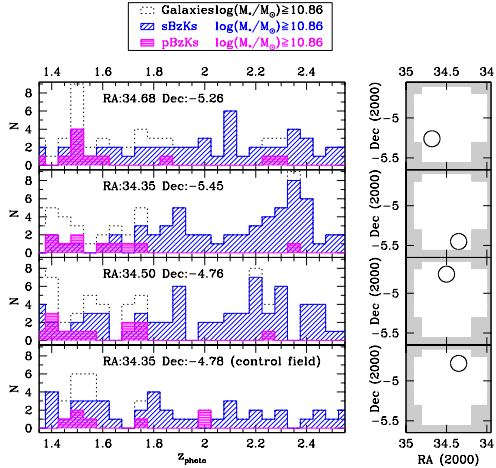

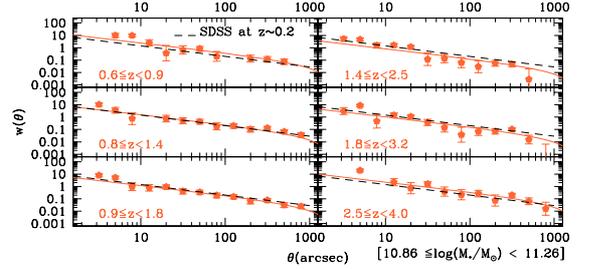

FIG. 10.— Angular correlation function of galaxies selected with $10.86 \leq \log(M_*/M_\odot) < 11.26$ at each redshift interval. Dashed lines (identical among all the six panels) show the correlation function derived from SDSS for the magnitude interval of $-22 < M_{r^*} < -21$ (Budavári et al. 2003). The SDSS galaxy sample has a median redshift of $z = 0.2$.

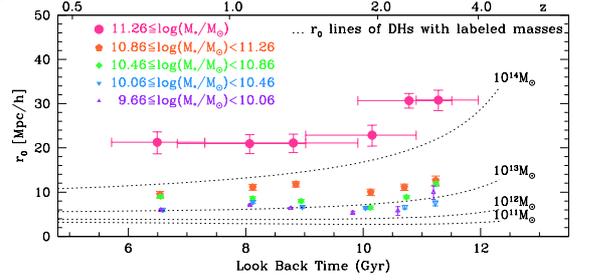

FIG. 11.— Correlation length, $r_0$ of galaxies, as a function of look back time for five mass intervals. These mass and time intervals are the same as the values listed in Table 2. Filled symbols show our data with the M/L completeness level larger than 95%. Dotted lines show the correlation lengths of dark haloes with different typical masses as labeled, based on the work of Mo & White (2002).

FIG. 9.— The photo-$z$ distributions for sub-samples of the massive galaxies with $log(M_*/M_\odot)$ at $1.4 \leq z_{photo} < 2.5$. The galaxies within the 0.1 deg radius circle of three over density regions and a control field (lowermost panel) are plotted. The right panels show the sky positions of each field.

| Redshift | | $0.6 \leq z < 0.9$ | | | $0.8 \leq z < 1.4$ | |
|---|---|---|---|---|---|---|
| Stellar mass | N | $A(arcsec)$ | $r_0[h^{-1}Mpc]$ | N | $A(arcsec)$ | $r_0[h^{-1}Mpc]$ |
| $10^{11.26}M_\odot \leq M_*$ | 76 | $73.5 \pm 15.4$ | $21.3^{+2.4}_{-2.6}$ | 153 | $32.3 \pm 5.9$ | $21.0^{+2.1}_{-2.2}$ |
| $10^{10.86}M_\odot \leq M_* < 10^{11.26}M_\odot$ | 409 | $15.6 \pm 2.2$ | $9.5^{+0.7}_{-0.7}$ | 775 | $10.0 \pm 1.0$ | $11.1^{+0.6}_{-0.7}$ |
| $10^{10.46}M_\odot \leq M_* < 10^{10.86}M_\odot$ | 729 | $14.0 \pm 1.1$ | $9.0^{+0.4}_{-0.4}$ | 1295 | $6.4 \pm 0.6$ | $8.5^{+0.5}_{-0.5}$ |
| $10^{10.06}M_\odot \leq M_* < 10^{10.46}M_\odot$ | 1006 | $7.0 \pm 0.8$ | $6.0^{+0.4}_{-0.4}$ | 1754 | $5.8 \pm 0.5$ | $7.8^{+0.2}_{-0.2}$ |
| $10^{9.66}M_\odot \leq M_* < 10^{10.06}M_\odot$ | 1595 | $6.9 \pm 0.5$ | $6.0^{+0.2}_{-0.3}$ | 2869 | $4.8 \pm 0.3$ | $7.1^{+0.2}_{-0.2}$ |

| Redshift | | $0.9 \leq z < 1.8$ | | | $1.4 \leq z < 2.5$ | |
|---|---|---|---|---|---|---|
| Stellar mass | N | $A(arcsec)$ | $r_0[h^{-1}Mpc]$ | N | $A(arcsec)$ | $r_0[h^{-1}Mpc]$ |
| $10^{11.26}M_\odot \leq M_*$ | 217 | $22.3 \pm 3.9$ | $21.1^{+2.0}_{-2.2}$ | 207 | $24.6 \pm 4.6$ | $22.9^{+2.3}_{-2.5}$ |
| $10^{10.86}M_\odot \leq M_* < 10^{11.26}M_\odot$ | 1142 | $7.6 \pm 0.7$ | $11.8^{+0.6}_{-0.6}$ | 1222 | $5.3 \pm 0.7$ | $10.0^{+0.7}_{-0.7}$ |
| $10^{10.46}M_\odot \leq M_* < 10^{10.86}M_\odot$ | 2268 | $4.1 \pm 0.4$ | $8.0^{+0.4}_{-0.4}$ | 2814 | $2.5 \pm 0.3$ | $6.5^{+0.4}_{-0.4}$ |
| $10^{10.06}M_\odot \leq M_* < 10^{10.46}M_\odot$ | 3282 | $3.2 \pm 0.2$ | $6.7^{+0.3}_{-0.3}$ | 3816 | $2.5 \pm 0.2$ | $6.4^{+0.3}_{-0.3}$ |
| $10^{9.66}M_\odot \leq M_* < 10^{10.06}M_\odot$ | 4202 | $2.8 \pm 0.2$ | $6.3^{+0.2}_{-0.2}$ | 2876 | $2.3 \pm 0.4$ | $5.3^{+0.4}_{-0.4}$ |

| Redshift | | $1.8 \leq z < 3.2$ | | | $2.5 \leq z < 4.0$ | |
|---|---|---|---|---|---|---|
| Stellar mass | N | $A(arcsec)$ | $r_0[h^{-1}Mpc]$ | N | $A(arcsec)$ | $r_0[h^{-1}Mpc]$ |
| $10^{11.26}M_\odot \leq M_*$ | 196 | $42.0 \pm 4.1$ | $30.7^{+1.6}_{-1.7}$ | 95 | $68.0 \pm 9.2$ | $30.8^{+2.3}_{-2.4}$ |
| $10^{10.86}M_\odot \leq M_* < 10^{11.26}M_\odot$ | 1039 | $6.7 \pm 0.4$ | $11.1^{+0.7}_{-0.7}$ | 367 | $14.4 \pm 2.1$ | $12.7^{+1.0}_{-1.1}$ |
| $10^{10.46}M_\odot \leq M_* < 10^{10.86}M_\odot$ | 2334 | $4.5 \pm 0.3$ | $8.9^{+0.4}_{-0.4}$ | 861 | $12.1 \pm 0.9$ | $11.9^{+0.5}_{-0.5}$ |
| $10^{10.06}M_\odot \leq M_* < 10^{10.46}M_\odot$ | 2848 | $2.6 \pm 0.3$ | $6.6^{+0.4}_{-0.4}$ | 1021 | $5.7 \pm 0.8$ | $7.5^{+0.6}_{-0.6}$ |
| $10^{9.66}M_\odot \leq M_* < 10^{10.06}M_\odot$ | 1373 | $2.2 \pm 0.7$ | $5.8^{+0.9}_{-1.1}$ | 315 | $10.7 \pm 2.9$ | $10.0^{+1.4}_{-1.6}$ |

TABLE 2
THE NUMBER COUNT, THE AMPLITUDE OF ACFs AND THE SPATIAL CORRELATION LENGTH IN EACH STELLAR MASS BIN FOR EACH REDSHIFT RANGE.

## 5. DISCUSSION

### 5.1. *Evolution of the stellar masses in galaxies*

Our wide and deep multi-color data set allows us to investigate the mass-dependent galaxy evolution with the spatial distributions. We construct the galaxy stellar mass functions (GSMFs) over a redshift range $0 \leq z < 4$ and derive the stellar mass density (SMD). We compare our GSMFs with those in previous studies (Drory et al. 2005; Fontana et al. 2006; Pérez-González et al. 2008; Elsner et al. 2008; Ilbert et al. 2010), which cover the redshift ranges overlapping with our study (Figure 6). Note, that most of these previous studies did not cover wide enough areas to statistically study massive galaxies at low redshift, nor they cover deep enough depth to study low-mass galaxies at high redshift. Compared to these studies, we are able to estimate the stellar mass functions more accurately over a wider redshift range, and also we are able to investigate properties of the $K$-selected galaxies to much lower-mass galaxies ($M_* < 10^{10.5} M_\odot$) at higher redshift (up to $3 \leq z < 4$).

From Figure 6, we can see that in our sample the number density of galaxies at the higher redshift range gets lower, as is consistent with other studies. Our rate of decrease in the number density of high-mass galaxies ($M_* > 10^{11.5} M_\odot$) is somehow faster at $z \gtrsim 2.0$, compared with the number density at the local Universe obtained by Cole et al.(2001).

Figure 7 shows the cosmological evolution of the stellar mass density (SMD). It shows that at the high redshift ($1.0 < z < 4.0$) the SMD evolved rapidly, then it slows down at the lower redshift ($z < 1.0$). The lower panel of Figure 7 shows the SMD evolution of the massive galaxies ($\log(M_*/M_\odot) \geq 10.86$). Although the uncertainty in the evolution of massive galaxies is relatively large, because of the small number of galaxies, the figure shows the evolution of SMD slowed down considerably at $z \lesssim 2$. This is consistent with Fontana et al. (2006) which suggested that about a half of massive galaxies have already been formed by $z \sim 1.5$.

In Figure 12, we plot the evolution of the number density (ND: left panel) and the SMD (right panel) for different mass ranges. The ND and the SMD are normalized by $\rho_N(M, z_{0.0-0.6})$ and $\rho_*(M, z_{0.0-0.6})$, respectively. As is suggested by Kodama et al. (2004), the ND of galaxies with smaller stellar masses increases later than larger stellar mass galaxies. From the right panel of Figure 12, it can be seen that the peak of the increase in the SMD for the larger stellar mass galaxies occurred at the higher redshift ($2 < z < 4$) than that of the smaller mass ones. This result suggests that the star formation in the massive galaxies is more active at the high redshift ($z > 2$) than at the lower redshift. This tendency is known as "downsizing" in which the active star-forming epoch of the massive galaxies is earlier than that of the smaller stellar mass galaxies.

Of course, this does not mean that all the smaller stellar mass galaxies are formed at low redshift. In fact, Figure 12 shows that there are a significant number of the smaller stellar mass galaxies ($10^{10.46} M_\odot \leq M_* < 10^{10.86} M_\odot$) already formed at $z > 2$. Our results show the fact that smaller stellar mass galaxies did not form later than massive galaxies, instead, the star formation activity of massive galaxies have subsided earlier than that of the smaller stellar mass galaxies. It is noteworthy that these high redshift smaller stellar mass galaxies outnumber the low redshift counterparts. This may suggest some of these high redshift smaller stellar mass galaxies may be merged and evolve into the present-day massive galaxies.

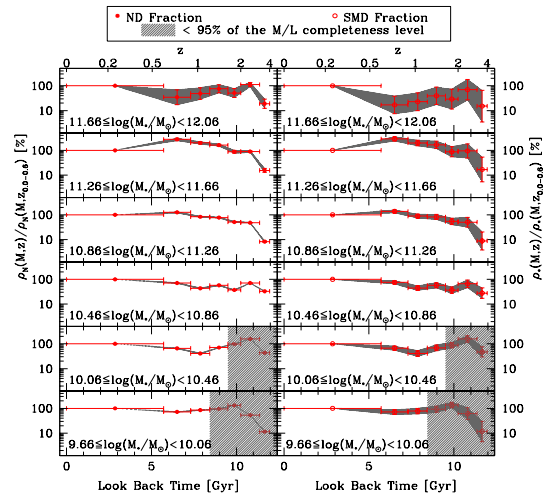

Fig. 12.— Dependence of the number density (ND) fractions (the left panels) and SMD fractions (the right panels) on the look back time. Filled circles show the ND fractions and open circles show the SMD fractions at each redshift interval within each stellar mass range as labeled. The gray hatted area show the range of the M/L completeness level less than 95%.

### 5.2. *Clustering as a function of redshift*

To investigate the clustering properties of galaxies, we measure the angular correlation function $\omega(\theta)$ and derive a spatial (comoving) correlation length $r_0$ as a function of redshift for various galaxy stellar mass ranges. Table 2 shows amplitudes of the ACFs, from which we see that the larger stellar mass galaxies have larger ACF amplitudes up to $z = 4$. This result indicates that the clustering of massive galaxies is stronger than that of the smaller stellar mass galaxies throughout our studied redshift ranges.

It is known that a large fraction of early-type galaxies are preferentially located in the clusters of galaxies. It is reported that rich clusters of galaxies in the local Universe typically reside in massive DHs with the mass of larger than $10^{14} M_\odot$. Figure 11 shows the evolution of the correlation length for various different galaxy mass ranges. From which, we can see the $r_0$ decreases rapidly from $z = 4$ to $z = 2$, and then the rate of decreasing $r_0$ becomes smaller to $z < 2$. Also, from Figure 11, it can be seen that the massive DHs host the most massive galaxies. On the other hand, the smaller stellar mass galaxies ($M_* < 10^{11.26} M_\odot$) are found in somewhat lower-mass DHs.

In Figure 11, we see that for the galaxies with $10^{10.86} M_\odot \leq M_* < 10^{11.26} M_\odot$ the $r_0$ value shifts at around $z = 1.8$. According to the standard CDM model prediction (Mo & White 2002), the typical masses of the DHs which host these galaxies increases from $2 \times 10^{13} M_\odot$ at $z > 1.8$ to $6 \times 10^{13} M_\odot$ at $z < 1.8$. The shifts in $r_0$ may imply this effect, although the prediction assumes a one-to-one correspondence between galaxies and haloes, and a more accurate comparison would require a reasonable halo occupation model. Our result shows that $r_0$ in the most massive sub-sample ($M_* > 10^{11.26} M_\odot$) is much larger than that of the smaller stellar mass galaxies. We find that those galaxies with the largest stellar masses are hosted by massive DHs ($1 - 5 \times 10^{14} M_\odot$). In addition, the typical mass of the DHs for the most massive galaxies at $z \sim 1$ should be larger than those at $z \sim 2$. The N-body simulation of $\Lambda$CDM structure formation model by Zhao et al. (2009) predicts that the mass of dark matter halos derived from the median mass accretion histories changes from $2.7 \times 10^{13} M_\odot$ at $z = 2$ to $5 \times 10^{14} M_\odot$ at the local Universe. These facts suggest that the most massive galaxies ($M_* > 10^{11.26} M_\odot$) at each redshift range may be the progenitors of the massive galaxies seen in the clusters in the local Universe. On the other hand, the change in $r_0$ of the intermediate-mass galaxies ($10^{10.46} M_\odot \leq M_* < 10^{10.86} M_\odot$) may not correlate with that of the DHs. The typical mass of the DHs hosting the intermediate-mass galaxies increase from smaller mass ($\sim 10^{12.8} M_\odot$) at $z \sim 2$ to larger mass ($\sim 10^{13.5} M_\odot$) at $z \sim 1$. For the less massive galaxies, however, this tendency is not clearly seen.

Figure 13 compares $r_0$ for our sample galaxies with those of various types of galaxy populations from previous studies. For $z \lesssim 2.5$, the value $r_0$ of massive galaxies is comparable to that of the local clusters of galaxies. The $r_0$ values of galaxies with $10^{10.86} M_\odot \leq M_* < 10^{11.26} M_\odot$ at $z \sim 1$ is as large as that of EROs. This agrees with the result that EROs are massive galaxies. Also, the DRGs, as well as the EROs, are candidate progenitors of the present-day massive galaxies. The $r_0$ values for the DRGs may vary, although those are not lower than the LBGs values, because it is known that the DRGs have an overlap with $sBzK$s at $1.4 \leq z \leq 2.5$ (92% in Grazian et al. 2007). The DRG population may include a considerable number of galaxies which are on the way to growing up to be more massive galaxies.

The observed change in $r_0$ of the intermediate-mass galaxies does not agree with the model prediction for the given DH masses. In contrast, it is seen that the change in $r_0$ values of the small stellar mass galaxies ($M_* < 10^{10.5} M_\odot$) shows a good agreement with the model. However, at $z \sim 1.0$, there is a possible increase in $r_0$ of the small stellar mass galaxies. The $r_0$ values for our low-mass-end sample and the LBGs, whose stellar masses may be low ($M_* = 10^{8-10} M_\odot$, e.g., Pentericci et al. 2008), are more or less constant over the redshift.

The $r_0$ values of the $sBzK$s and of the $pBzK$s in this study are estimated by using all the sample galaxies with various masses (filled symbols enclosed by the same shapes). There is a tendency for the $r_0$ of the $pBzK$s to be larger than that of the $sBzK$s, when we compare the two values for the same stellar mass range. This is consistent with the model prediction in which the $pBzK$s are hosted in more massive DHs (Hartley et al. 2008; Blanc et al. 2008) than the $sBzK$s. Our result shows that the correlation length $r_0$ of the whole $pBzK$ sample is larger than that of the whole $sBzK$ sample. The $pBzK$ galaxies, which are the galaxies with little star formation activity, have larger stellar masses, while the star-forming $sBzK$ galaxies have a wide range of stellar masses.

Since 77% of the massive galaxies ($M_* \geq 10^{11.26} M_\odot$) at $1.4 \leq z < 2.5$ in our sample satisfy the $sBzK$s criteria, the clustering structures seen for these massive galaxies at this epoch should be dominated by the $sBzK$s. However, $r_0$ of the $pBzK$s is relatively large compared to that of the $sBzK$s. These results may suggest that the structures seen for the $pBzK$s were formed earlier and were preserving the information of spatial distribution of the massive DHs.

When the UKIDSS/UDS observations are progressed, we should be able to use a deeper near-IR data set and discuss the clustering property including the lower-mass galaxies, and compare directly the mass-dependent clustering property of our sample galaxies with that of the LBGs at $z > 2.5$.

### 5.3. *Spatial distributions of galaxies at $1.4 \leq z < 2.5$*

It is known that the space density of the local clusters of galaxies is of the order of $10^{-6} - 10^{-7} h^{-3} Mpc^{-3}$ (Bramel et al. 2000). Since the survey volume of the SXDS/UDS field at $1.4 \leq z < 2.5$ is $6.2 \times 10^6 Mpc^3$, it may be possible to find some proto-cluster candidates in our field. Figure 8 shows the mass-density map of massive galaxies ($M_* \gtrsim 10^{11.26} M_\odot$) in the redshift range $1.4 \leq z < 2.5$. We see several high mass-density regions with sizes of $\sim 8 \times 8 Mpc/h^2$, which are close to typical sizes of rich clusters of galaxies in the local Universe. As described in Section 5.2, 77% of the massive galaxies at $1.4 \leq z < 2.5$ are the $sBzK$s. Thus, the star formation of most massive galaxies at this epoch is very active. Also, we see the tendency that these galaxies are found in high mass density regions. On the other hand, no clear concentration of the massive $pBzK$s is seen in the photo-$z$ distributions for sub samples of the massive galaxies at $1.4 \leq z_{photo} < 2.5$ (Figure 9). At $1.4 \leq z < 2.5$, the $pBzK$s are older than the $sBzK$s. Since $r_0$ of the $pBzK$s ($14.7^{+1.1}_{-1.1}$) is larger than that of the $sBzK$s ($5.3^{+0.1}_{-0.1}$), the $pBzK$s must be residing in the larger mass DHs. We find that the average stellar mass of the $pBzK$s at $1.4 \leq z < 2.5$ is $M_* = 10^{10.8} M_\odot$. Although this average value corresponds to that of the intermediate-mass galaxies, $r_0$ of the $pBzK$s is much larger than that for the whole galaxy populations in the same mass range.

Since larger stellar mass galaxies are hosted by larger mass DHs, we can see that the passively-evolving galaxies (i.e. $pBzK$s) tend to belong to the larger mass DHs at $1.4 \leq z < 2.5$. This result suggests that the galaxies which are hosted by massive DHs may have started their star formation at earlier epochs and stopped or considerably subsided their star formation activities by $z \sim 2$, which implies the dependency of star formation on the host DH masses. The difference in the 2-D distribution of these massive $sBzK$ and $pBzK$ galaxies supports that the formation epoch of the $pBzK$ galaxies is earlier than that of the $sBzK$ galaxies. This suggests that these $pBzK$ galaxies were formed under stronger clustering environments. Although our sample has only a small number of the massive $pBzK$s, it shows no concentration as described above, and it has no clear correlation with the redshift distribution of the $sBzK$s.

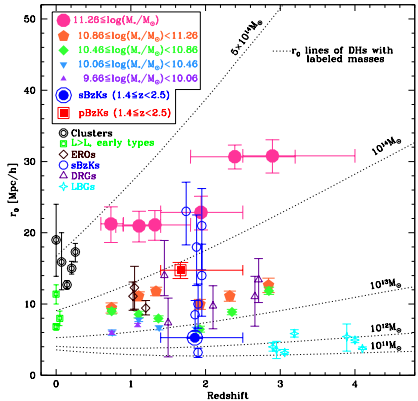

Fig. 13.— The redshift evolution of correlation lengths of various types of objects. Filled symbols show our result. The correlation lengths of our $sBzK$s and $pBzK$s at $1.4 \leq z < 2.5$ are plotted as the blue open circle containing filled circle and the red open square containing filled square. Dotted lines show the correlation lengths of dark haloes with four different typical masses, which are same as values on Figure 11. Hollow symbols show correlation lengths of the different populations of objects (Double black circles: present-day clusters of galaxies from Bahcall et al. [2003]. Double light green circles: present-day luminous early-type galaxies and Brown diamonds: EROs from Overzier et al. [2003] and see references therein. Purple triangles: DRGs, Blue circles: $sBzK$s with various $K$ magnitude range and Cyan crosses: LBGs from Hayashi et al. [2007] and see references therein. ).

## 6. SUMMARY

In this paper, we investigate the mass-dependent evolution of galaxies in the redshift range $0.6 \leq z < 4$ using the wide and deep optical and near-infrared imaging surveys, the SXDS DR1 and the UKIDSS/UDS DR1. From which, we obtain a sub-sample of 55,174 $K$-selected galaxies, which are detected in at least 6 optical to near-infrared bands (i.e. 5 data points in addition to the $K$-band), to a limiting magnitude of $K_{Vega} = 21.55$ over an area of 0.63 deg$^2$. Then, further masking out low quality data regions, we derive the photometric redshifts and stellar masses of 48,569 galaxies over an area of 0.53 deg$^2$. Because of the wide survey area we have, the clustering properties derived are relatively free from the field-to-field variation compared to the previous studies of comparable depth. This enables us to investigate not only the evolution of stellar mass of galaxies, but also the clustering evolution of galaxies as a function of the stellar mass up to the redshift $z = 4$.

Our results are as follows:

- The $K$-selected GSMFs in the redshift range $0 \leq z < 4$ obtained for the SXDS/UDS field are consistent with the previous studies except for the high-mass end, which our data show systematically lower stellar masses.

- The change in the evolutionary track of the SMD of massive galaxies ($M_* \geq 10^{10.86} M_\odot$) is getting considerably smaller at $z \lesssim 2.0$, which indicates that the star formation activity of massive galaxies is subsided at around this redshift.

- A considerable number of low mass galaxies ($M_* < 10^{10.5} M_\odot$) are found at high redshift, $3 \leq z < 4$.

- More massive galaxies have stronger clustering through out the redshift up to $z = 4$.

- The largest mass dark haloes host the largest stellar mass galaxies.

- At the lower redshift, $z < 2$, the clustering strength of the largest stellar mass galaxies ($M_* \geq 10^{11.3} M_\odot$) is weaker than that of the higher redshift, $z > 2$.

- For the intermediate-mass galaxies ($10^{10.46} M_\odot \leq M_* < 10^{10.86} M_\odot$), the change in clustering strength with the redshift depends weakly on the galaxy mass.

- At the redshift range $1.4 \leq z < 2.5$, several high mass-density regions are identified, which contain the most massive $sBzK$ galaxies ($M_* > 10^{11.3} M_\odot$). These could be the proto-clusters which evolve into the present-day rich clusters of galaxies.

- No obvious relationship is found between the redshift distributions of the massive $sBzK$s and the massive $pBzK$s in our sample.

- At the redshift range $1.4 \leq z < 2.5$, the correlation length $r_0$ of the $pBzK$s is much larger than that of the $sBzK$s. This can be explained if the $pBzK$s at $1.4 \leq z < 2.5$ are the intermediate-mass galaxies formed at the higher redshift and star formation has been terminated by $z \sim 2$, so that they got imprinted the clustering properties of the earlier epoch.

We thank the Subaru telescope and the United Kingdom Infrared Telescope for the assistance in obtaining the high-quality data set.